\documentclass[submission,copyright,creativecommons]{eptcs}

\providecommand{\event}{F-IDE 2016} 

\usepackage{breakurl}             
\usepackage{underscore}           
\usepackage{amssymb}
\usepackage{amsmath}
\usepackage{graphicx}
\usepackage{tabularx,multicol,multirow}
\usepackage{booktabs}
\usepackage{url}
\usepackage{pifont}
\usepackage[]{algorithm2e}
\usepackage{tikz}

\newcommand{\upbar}{\tikz[overlay] \draw (-.5em,1em)--(-.5em,0em);}
\newcommand{\downbar}{\tikz[overlay] \draw (-.5em,.8em)--(-.5em,-1em);}
\urldef{\mailsa}\path|{ahealy, rosemary, jpower}@cs.nuim.ie|
\newcommand{\why}{\textsf{Why3}}%

\usetikzlibrary{arrows,shapes,positioning,shadows,trees}

\tikzset{
  basic/.style  = {draw, text width=2cm, drop shadow, font=\sffamily, rectangle},
  root/.style   = {basic, rounded corners=6pt, thin, align=center, fill=cyan!50},
  level 2/.style = {basic, rounded corners=6pt, thin,align=center, fill=cyan!50, text width=6em},
  level 3/.style = {basic, thin, align=left, fill=pink!60, text width=3.5em}
}

\begin{document}


\title{Predicting SMT Solver Performance for Software Verification}


%
%
\author{Andrew Healy\quad\quad Rosemary Monahan\quad\quad  James F. Power
\institute{Dept. of Computer Science, Maynooth University, Maynooth, Ireland}
\email{ahealy@cs.nuim.ie\quad\quad rosemary@cs.nuim.ie\quad\quad jpower@cs.nuim.ie}
}
\def\titlerunning{Predicting SMT Solver Performance for Software Verification}
\def\authorrunning{A. Healy, R. Monahan \& J. F. Power}



%
%

\maketitle

\begin{abstract}
The \textsf{Why3} IDE and verification system facilitates the use of a wide range of Satisfiability Modulo Theories (SMT) solvers through a driver-based architecture. We present \textsf{Where4}: a portfolio-based approach to discharge \textsf{Why3} proof obligations. We use data analysis and machine learning techniques on static metrics derived from program source code. Our approach benefits software engineers by providing a single utility to delegate proof obligations to the solvers most likely to return a useful result. It does this in a time-efficient way using existing \textsf{Why3} and solver installations --- without requiring low-level knowledge about SMT solver operation from the user.
\end{abstract}

\section{Introduction}

\label{sec:intro}

The formal verification of software generally requires a software engineer to use a system of tightly integrated components. Such systems typically consist of an IDE that can accommodate both the implementation of a program and the specification of its formal properties. 
These two aspects of the program are then typically translated into the logical constructs of an intermediate language, forming a series of goals which must be proved in order for the program to be fully verified. These goals (or ``proof obligations'') must be formatted for the system's general-purpose back-end solver. Examples of systems which follow this model are Spec\# \cite{spec} and Dafny \cite{Dafny} which use the Boogie \cite{Boogie} intermediate language and the Z3 \cite{Z3} SMT solver. 

\textsf{\textsf{Why3}} \cite{why:whereprograms} was developed as an attempt to make use of the wide spectrum of interactive and automated theorem proving tools and overcome the limitations of systems which rely on a single SMT solver. It provides a driver-based, extensible architecture to perform the necessary translations into the input formats of two dozen provers. With a wide choice of theorem-proving tools now available to the software engineer, the question of choosing the most appropriate tool for the task at hand becomes important. It is this question that \textsf{Where4} answers.

As motivation for our approach, Table \ref{table:avgtimes} presents the results from running the \textsf{Why3} tool over the example programs included in the \textsf{Why3} distribution (version 0.87.1), using eight SMT solvers at the back-end.  
Each \textsf{Why3} file contains a number of theories requiring proof, and these in turn are  broken down into a number of goals for the SMT solver; for the data in Table \ref{table:avgtimes} we had 128 example programs, generating 289 theories, in turn generating 1048 goals.  In Table \ref{table:avgtimes} each row presents the data for a single SMT solver, and the three main data columns give data totalled on a per-file, per-theory and per-goal basis. Each of these three columns is further broken down to show the number of programs/theories/goals that were successfully solved, their percentage of the total, and the average time taken in seconds for each solver to return such a result. Program verification by modularisation construct is particularly relevant to the use of \textsf{Why3} on the command line as opposed to through the IDE.

Table \ref{table:avgtimes} also has a row for an imaginary ``theoretical'' solver, \textsf{Choose Single}, which corresponds to choosing the best (fastest) solver for each individual program, theory or goal.  This solver performs significantly better than any individual solver, and gives an indication of the maximum improvement that could be achieved \textit{if it was possible to predict in advance which solver was the best for a given program, theory or goal}.  In general, the method of choosing from a range of solvers on an individual goal basis is called \textit{portfolio-solving}. This technique has been successfully implemented in the SAT solver domain by SATzilla \cite{Satzilla} and for model-checkers \cite{DPVZ15:CAV}\cite{MUX}. \textsf{Why3} presents a unique opportunity to use a common input language to develop a portfolio SMT solver specifically designed for software verification.  

The main contributions of this paper are:
\begin{enumerate}
\item The design  and implementation of our portfolio solver, \textsf{Where4}, which uses supervised machine learning to predict the best solver to use based on metrics collected from goals.
\item The integration of \textsf{Where4} into the user's existing \textsf{Why3} work-flow by imitating the behaviour of an orthodox SMT solver.
\item A set of metrics to characterise \textsf{Why3} goal formulae.
\item Statistics on the performance of eight SMT solvers using a dataset of 1048 \textsf{Why3} goals.

\end{enumerate}

\newcolumntype{Y}{>{\raggedleft\arraybackslash}X} 
\begin{table}
\caption{Results of running 8 solvers on the example \textsf{Why3} programs with a timeout value of 10 seconds. In total our dataset contained 128 files, which generated 289 theories, which in turn generated 1048 goals.  Also included is a theoretical solver  \textsf{Choose Single}, which always returns the best answer in the fastest time.}
\begin{tabularx}{\textwidth}{@{}l|YYY|YYY|YYY@{}}
\toprule
{} & \multicolumn{3}{c|}{\textbf{File}} & \multicolumn{3}{c|}{\textbf{Theory}} & \multicolumn{3}{c}{\textbf{Goal}} \\
{} & \# proved & \% proved & Avg time & \# proved & \% proved & Avg time & \# proved & \% proved & Avg time \\
\midrule
\textsf{Choose Single} & \textbf{48} & \textbf{37.5\%} & \textbf{1.90} & \textbf{190} & \textbf{63.8\%} & \textbf{1.03} & \textbf{837} & \textbf{79.9\%} & \textbf{0.42} \\
\textbf{Alt-Ergo-0.95.2} & 25 & 19.5\% & 1.45 & 118 & 39.6\%& 0.77 & 568 & 54.2\% & 0.54 \\ 
\textbf{Alt-Ergo-1.01} & 34 & 26.6\% & 1.70 & 142 & 47.7\% & 0.79 & 632 & 60.3\% & 0.48 \\ 
\textbf{CVC3} & 19 & 14.8\% & 1.06 & 128 & 43.0\% & 0.65 & 597 & 57.0\% & 0.49 \\ 
\textbf{CVC4} & 19  & 14.8\% & 1.09 & 117 & 39.3\% & 0.51 & 612 & 58.4\% & 0.37 \\ 
\textbf{veriT} & 5 & 4.0\% & 0.12 & 79 & 26.5\% & 0.20 & 333 & 31.8\% & 0.26 \\ 
\textbf{Yices} & 14 & 10.9\% & 0.53 & 102 & 34.2\% & 0.22 & 368 & 35.1\% & 0.22 \\ 
\textbf{Z3-4.3.2} & 25 & 19.5\% & 0.56 & 128 & 43.0\% & 0.36 & 488 & 46.6\% & 0.38 \\ 
\textbf{Z3-4.4.1} & 26 & 20.3\% & 0.58 & 130 & 43.6\% & 0.40 & 581 & 55.4\% & 0.35 \\ 
\bottomrule
\end{tabularx}
\label{table:avgtimes}
\end{table}
Section \ref{sec:overview} describes how the data was gathered and discusses issues around the accurate measurement of results and timings. A comparison of prediction models forms the basis of Section \ref{sec:predselection} where a number of evaluation metrics are introduced. The \textsf{Where4} tool is compared to a range of SMT tools and strategies in Section \ref{sec:eval}. The remaining sections present a review of additional related work and a summary of our conclusions. 

\section{System Overview and Data Preparation}
\label{sec:overview}

Due to the diverse range of input languages used by software verification systems, a standardised benchmark repository of verification programs does not yet exist \cite{Dagstuhl}.
For our study we chose the 128 example programs included in the \textsf{Why3} distribution (version 0.87.1) as our corpus for training and testing purposes. The programs in this repository are written in WhyML, a dialect of ML with added specification syntax and verified libraries. Many of the programs are solutions to problems posed at software verification competitions such as VerifyThis \cite{verifythis}, VSTTE \cite{Klebanov2011} and COST \cite{bormer:hal-00789525}. Other programs are implementations of benchmarks proposed by the VACID-0 \cite{Leino10vacid-0:verification} initiative.   It is our assumption that these programs are a representative software verification workload. Alternatives to this dataset are discussed in Section \ref{sec:related}. 
    
We used six current, general-purpose SMT solvers supported by \textsf{Why3}: Alt-Ergo \cite{AltErgo} versions 0.95.2 and 1.01, CVC3 \cite{CVC3} ver. 2.4.1, CVC4 \cite{CVC4} ver. 1.4, veriT \cite{veriT}, ver. 201506\footnote{The most recent version - 201506 - is not officially supported by \textsf{Why3} but is the only version available}, Yices \cite{Yices} ver. 1.0.38\footnote{We did not use Yices2 as its lack of support for quantifiers makes it unsuitable for software verification}, and Z3 \cite{Z3} ver. 4.3.2 and 4.4.1. We expanded the range of solvers to eight by recording the results for two of the most recent major versions of two popular solvers - Alt-Ergo and Z3.

\begin{figure}
\centering
\includegraphics[width=0.7\linewidth]{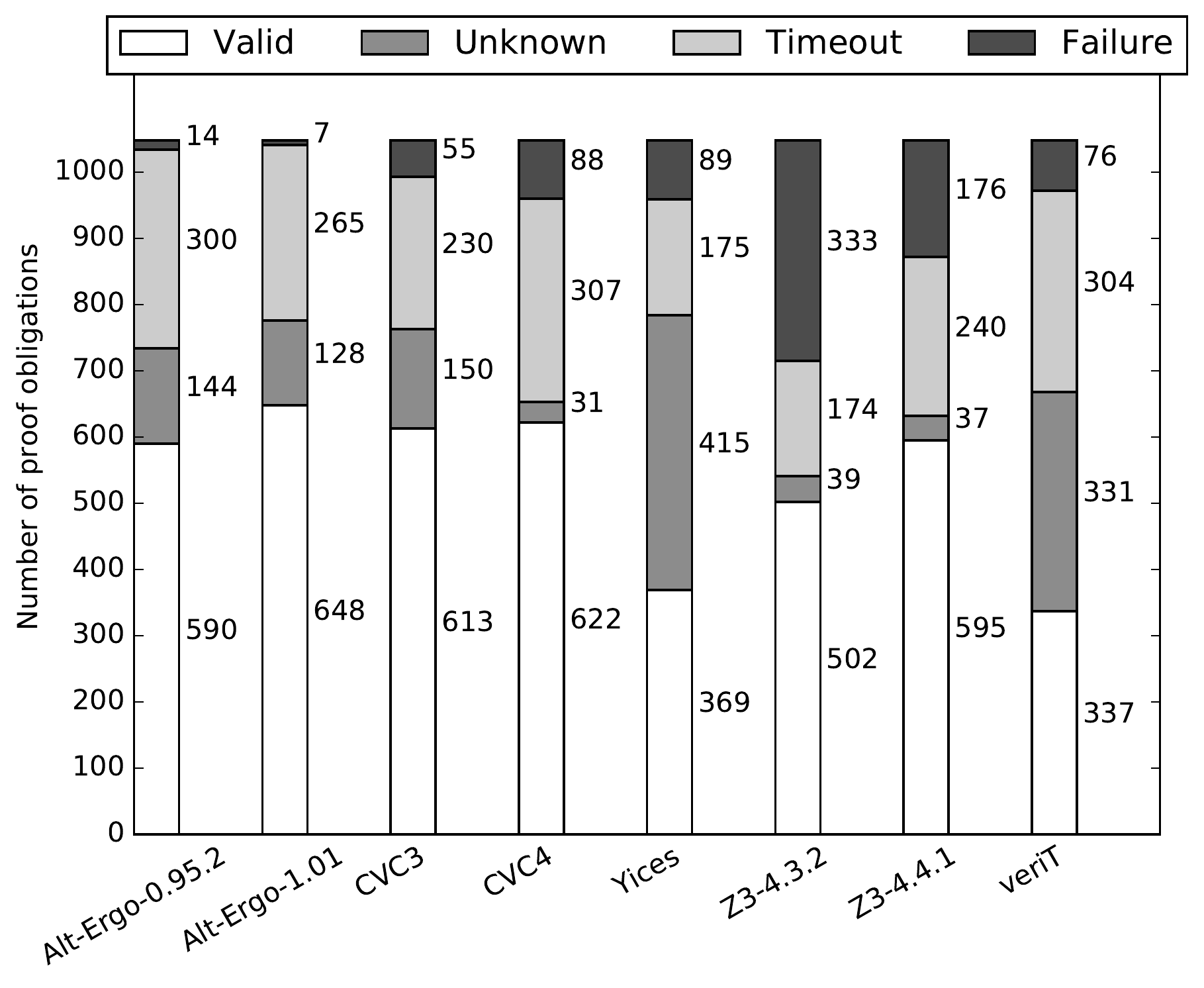}
\caption{The relative amount of \textit{Valid/Unknown/Timeout/Failure} answers from the eight SMT solvers (with a timeout of 60 seconds). Note that no tool returned an answer of \textit{Invalid} for any of the 1048 proof obligations.}
\label{fig:barcharts}
\end{figure}
    
When a solver is sent a goal by \textsf{Why3} it returns one of the five possible answers \textit{Valid},  \textit{Invalid},  \textit{Unknown},  \textit{Timeout} or  \textit{Failure}. As can be seen from Table \ref{table:avgtimes} and Fig. \ref{fig:barcharts}, not all goals can be proved Valid or Invalid. Such goals usually require the use of an interactive theorem prover to discharge goals that require reasoning by induction. Sometimes a splitting transformation needs to be applied to simplify the goals before they are sent to the solver. Our tool does not perform any transformations to goals other than those defined by the solver's \textsf{Why3} driver file. In other cases, more time or memory resources need to be allocated in order to return a conclusive result. We address the issue of resource allocation in Section \ref{sec:independant}.     

\subsection{Problem Quantification: predictor and response variables}

Two sets of data need to be gathered in supervised machine learning \cite{Mitchell}: the independent/predictor variables which are used as input for both training and testing phases, and the dependent/response variables which correspond to ground truths during training. Of the 128 programs in our dataset, 25\% were held back for system evaluation (Section \ref{sec:eval}). The remaining 75\% (corresponding to 96 WhyML programs, 768 goals) were used for training and 4-Fold cross-validation.

\subsubsection{Independent/Predictor Variables}
\label{sec:independant}
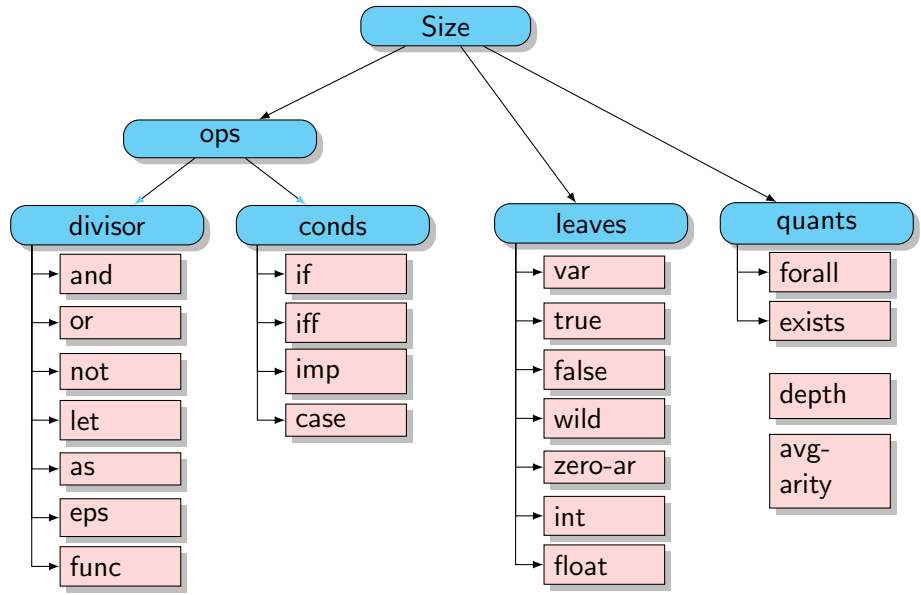
\begin{figure}
\centering
\begin{tikzpicture}[
  level 1/.style={sibling distance=30mm},
  edge from parent/.style={->,draw},
  >=latex]

\node[root]{Size}
  child {node[level 2] (c0) {ops}
    child {node[level 2, yshift=10pt] (c1) {divisor}}
    child {node[level 2, yshift=10pt] (c2) {conds}}
  }
  child {node[level 2, yshift=-32pt, xshift=55pt] (c3) {leaves}}
  child {node[level 2, yshift=-32pt, xshift=55pt] (c4) {quants}}
;

\begin{scope}[every node/.style={level 3}]
\node [below of = c1, xshift=5pt, yshift=10pt] (c11) {and};
\node [below of = c11, yshift=10pt] (c12) {or};
\node [below of = c12, yshift=10pt] (c13) {not};
\node [below of = c13, yshift=10pt] (c14) {let};
\node [below of = c14, yshift=10pt] (c15) {as};
\node [below of = c15, yshift=10pt] (c16) {eps};
\node [below of = c16, yshift=10pt] (c17) {func};

\node [below of = c2, xshift=5pt, yshift=10pt] (c21) {if};
\node [below of = c21, yshift=10pt] (c22) {iff};
\node [below of = c22, yshift=10pt] (c23) {imp};
\node [below of = c23, yshift=10pt] (c24) {case};

\node [below of = c3, xshift=5pt, yshift=10pt] (c31) {var};
\node [below of = c31, yshift=10pt] (c32) {true};
\node [below of = c32, yshift=10pt] (c33) {false};
\node [below of = c33, yshift=10pt] (c34) {wild};
\node [below of = c34, yshift=10pt] (c35) {zero-ar};
\node [below of = c35, yshift=10pt] (c36) {int};
\node [below of = c36, yshift=10pt] (c37) {float};

\node [below of = c4, xshift=5pt, yshift=10pt] (c41) {forall};
\node [below of = c41, yshift=10pt] (c42) {exists};

\node [below of = c42](c43) {depth};
\node [below of = c43](c44) {avg-arity};

\end{scope}
\foreach \value in {1,...,7}
  \draw[->] (c1.195) |- (c1\value.west);

\foreach \value in {1,...,4}
  \draw[->] (c2.195) |- (c2\value.west);

\foreach \value in {1,...,7}
  \draw[->] (c3.195) |- (c3\value.west);

\foreach \value in {1,...,2}
  \draw[->] (c4.195) |- (c4\value.west);

\end{tikzpicture}
\caption{Tree illustrating the Why syntactic features counted individually (\textit{pink nodes}) while traversing the AST. The rectangles represent individual measures, 
and the rounded blue nodes represent metrics that are the sum of their children in the tree.}
\label{fig:types}
\end{figure}

Fig. \ref{fig:types} lists the predictor variables that were used in our study.  All of these are (integer-valued) metrics that can be calculated by analysing a \textsf{Why3} proof obligation, and are similar to the \textit{Syntax} metadata category for proof obligations written in the TPTP format \cite{TPTP}.
To construct a feature vector from each task sent to the solvers, we traverse the abstract syntax tree (AST) for each goal and lemma, counting the number of each syntactic feature we find on the way. We 
focus on goals and lemmas as they produce proof obligations, with axioms and predicates providing a logical context.

Our feature extraction algorithm has similarities in this respect to the method used by \textsf{Why3} for computing goal ``shapes'' \cite{why:preserving}. These shape strings are used internally by \textsf{Why3} as an identifying fingerprint. Across proof sessions, their use can limit the amount of goals in a file which need to be re-proved.   

\subsubsection{Dependent/Response Variables}
\label{sec:dependant}

Our evaluation of the performance of a solver depends on two factors: the time taken to calculate that result, and whether or not the solver had actually proven the goal.

In order to accurately measure the time each solver takes to return an answer, we used a measurement framework specifically designed for use in competitive environments. The BenchExec \cite{benchexec} framework was developed by the organisers of the SVCOMP \cite{SVCOMP} software verification competition to reliably measure CPU time, wall-clock time and memory usage of software verification tools. We recorded the time spent on CPU by each SMT solver for each proof obligation. To account for random errors in measurement introduced at each execution, we used the methodology described by Lilja \cite{LiljaJ} to obtain an approximation of the true mean time. A 90\% confidence interval was used with an allowed error of $\pm$3.5\%.   

By inspecting our data, we saw that most \textit{Valid} and \textit{Invalid} answers returned very quickly, with \textit{Unknown} answers taking slightly longer, and \textit{Failure/Timeout} responses taking longest. We took the relative utility of responses to be $\lbrace Valid, Invalid\rbrace>Unknown>\lbrace Timeout,Failure\rbrace$ which can be read as ``it is better for a solver to return a \textit{Valid} response than \textit{Timeout}'', etc. A simple function allocates a cost to each solver $S$'s response to each goal $G$:
\[\small
	cost(S,G) = 
	\begin{cases}
		time_{S,G}, \text{ if answer}_{S,G} \in \lbrace Valid, Invalid \rbrace \\
		time_{S,G} + \text{timeout}, \text{ if answer}_{S,G} = Unknown \\
		time_{S,G} + (\text{timeout} \times 2), \text{ if answer}_{S,G} \in \lbrace Timeout, Failure \rbrace
	\end{cases}
\]

Thus, to penalise the solvers that return an \textit{Unknown} result, the timeout limit is added to the time taken, while solvers returning \textit{Timeout} or \textit{Failure} are further penalised by 
adding double the timeout limit to the time taken.
A response of \textit{Failure} refers to an error with the backend solver and usually means a required logical theory is not supported. 
This function ensures the best-performing solvers always have the lowest costs. A ranking of solvers for each goal in order of decreasing relevance is obtained by sorting the solvers by ascending cost.

Since our cost model depends on the time limit value chosen, we need to choose a value that does not favour any one solver.  To establish a realistic time limit value, we find each solver's ``Peter Principle Point'' \cite{Sutcliffe200139}. In resource allocation for theorem proving terms, this point can be defined as the time limit at which more resources will not lead to a significant increase in the number of goals the solver can prove. 

\begin{figure}
\centering
\includegraphics[width=0.7\linewidth]{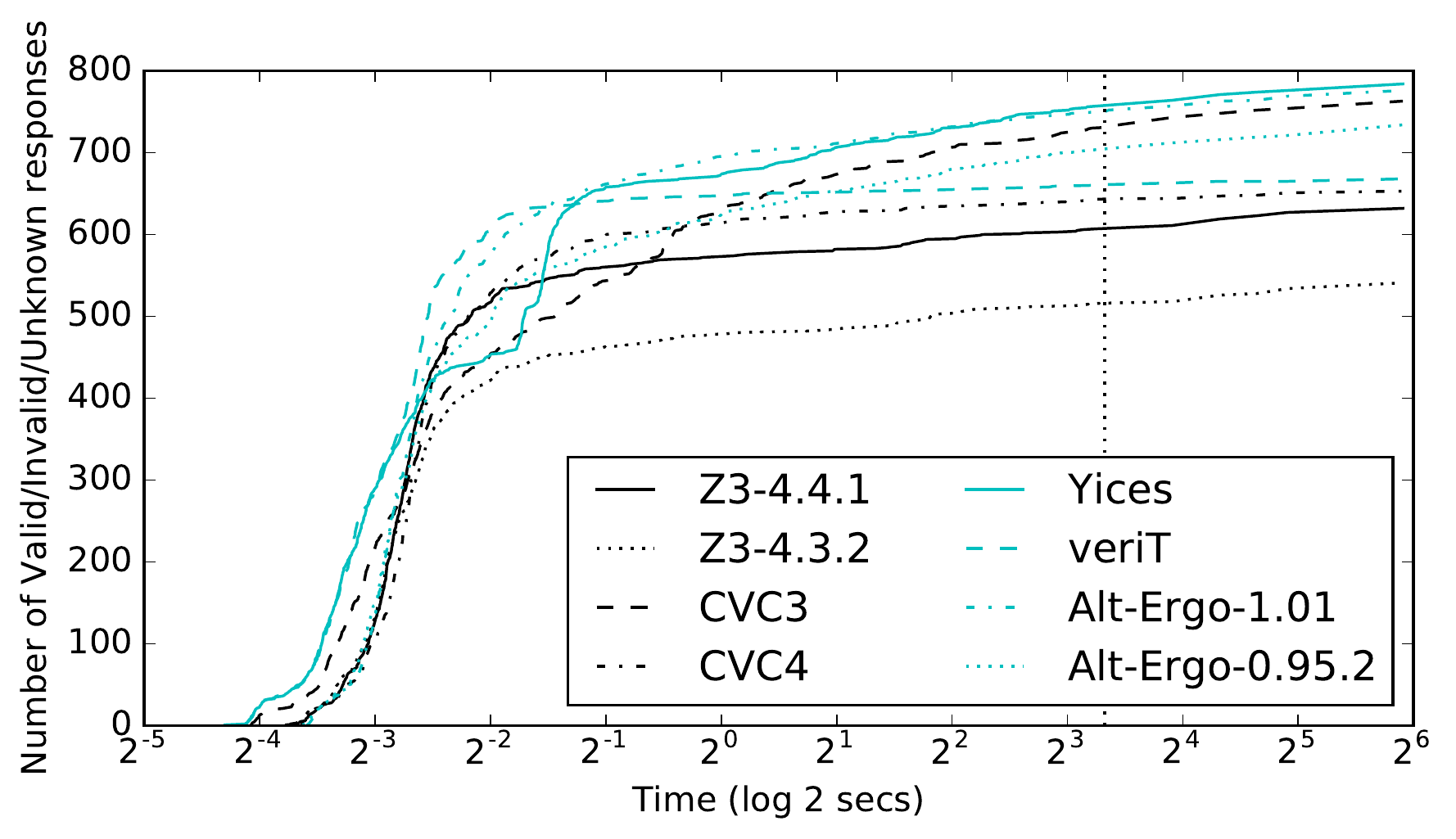}
\caption{The cumulative number of \textit{Valid/Invalid/Unknown} responses for each solver. The plot uses a logarithmic scale on the time axis for increased clarity at the low end of the scale. The chosen timeout limit of 10 secs (\textit{dotted vertical line}) includes 99\% of each solver's useful responses}
\label{fig:line-graph}
\end{figure}

Fig. \ref{fig:line-graph} shows the number of \textit{Valid/Invalid/Unknown} results for each prover when given a time limit of 60 seconds. 
This value was chosen as an upper limit, since a time limit value of 60 seconds is not realistic for most software verification scenarios.  \textsf{Why3}, for example, has a default time limit value of 5 seconds. 
From Fig. \ref{fig:line-graph} we can see that the vast majority of useful responses are returned very quickly. 

By satisfying ourselves with being able to record 99\% of the useful responses which would be returned after 60 seconds, a more reasonable threshold is obtained for each solver. This threshold ranges from 7.35 secs (veriT) to 9.69 secs (Z3-4.3.2). Thus we chose a value of 10 seconds as a representative, realistic time limit that gives each solver a fair opportunity to return decent results.     

\section{Choosing a prediction model}
\label{sec:predselection}
Given a \textsf{Why3} goal, a ranking of solvers can be obtained by sorting the cost for each solver. For unseen instances, two approaches to prediction can be used: (1) classification --- predicting the final ranking directly --- and (2) regression --- predicting each solver's score individually and deriving a ranking from these predictions.  With eight solvers, there are $8!$ possible rankings. Many of these rankings were observed very rarely or did not appear at all in the training data. Such an unbalanced dataset is not appropriate for accurate classification, leading us to pursue the regression approach.

Seven regression models were evaluated\footnote{We used the Python Sci-kit Learn \cite{sklearn} implementations of these models}: Linear Regression, Ridge Regression, K-Nearest Neighbours, Decision Trees, Random Forests (with and without discretisation) and the regression variant of Support Vector Machines. Table \ref{table:truncated} shows the results for some of the best-performing models. Most models were evaluated with and without a weighting function applied to the training samples. Weighting is standard practice in supervised machine learning: each sample's weight was defined as the standard deviation of solver costs. This function was designed to give more importance to instances where there was a large difference in performance among the solvers. 

Table \ref{table:truncated} also shows three theoretical strategies in order to provide bounds for the prediction models. \textit{Best} always chooses the best ranking of solvers and \textit{Worst} always chooses the worst ranking (which is the reverse ordering to \textit{Best}). \textit{Random} is the average result of choosing every permutation of the eight solvers for each instance in the training set.  We use this strategy to represent the user selecting SMT solvers at random without any consideration for goal characterisation or solver capabilities. 
A comparison to a \textit{fixed} ordering of solvers for each goal is not made as any such ordering would be arbitrarily determined. 

We note that the \textit{Best} theoretical strategy of Table \ref{table:truncated} is not directly comparable with the theoretical solver \textsf{Choose Single} from Table \ref{table:avgtimes}.  The two tables' average time columns are measuring different results: in contrast to \textsf{Choose Single}, \textit{Best} will call each solver in turn, as will all the other models in Table \ref{table:truncated}, until a result \textit{Valid/Invalid} is recorded (which it may never be).  Thus Table \ref{table:truncated}'s \textit{Time} column shows the average \textit{cumulative} time of each such sequence of calls, rather than the average time taken by the single \textit{best} solver called by \textsf{Choose Single}. 

\subsection{Evaluating the prediction models}

Table \ref{table:truncated}'s \textit{Time} column provides an overall estimate of the effectiveness of each prediction model.  We can see that the discretised Random Forest method provides the best overall results for the solvers, yielding an average time of 14.92 seconds.

The second numeric column of Table \ref{table:truncated} shows the 
Normalised Discounted Cumulative Gain (\textit{\textbf{nDCG}}), which is commonly used to evaluate the accuracy of rankings in the search engine and e-commerce recommender system domains \cite{NDCG}. Here, emphasis is placed on correctly predicting items higher in the ranking. For a general ranking of length $p$, it is formulated as:
\[\small
	nDCG_p = \frac{DCG_p}{IDCG_p}
    \quad\text{ where }\quad
    DCG_p = \sum_{i=1}^{p} \frac{2^{rel_i} - 1}{log_2(i+1)}
\]
Here $rel_i$ refers to the relevance of element $i$ with regard to a ground truth ranking, and we take each solver's relevance to be inversely proportional to its rank index.  In our case, $p = 8$ (the number of SMT solvers).  The $DCG_p$ is normalised by dividing it by the maximum (or \textit{idealised}) value for ranks of length $p$, denoted $IDCG_p$. As our solver rankings are permutations of the ground truth (making $nDCG$ values of 0 impossible), the values in Table \ref{table:truncated} are further normalised to the range [0..1] using the lower $nDCG$ bound for ranks of length 8 --- found empirically to be 0.4394. 

The third numeric column of Table \ref{table:truncated} shows
the $R^2$ score (or coefficient of determination), which is an established metric for evaluating how well regression models can predict the variance of dependent/response variables. The maximum $R^2$ score is 1 but the minimum can be negative. Note that the theoretical strategies return rankings rather than individual solver costs. For this reason, $R^2$ scores are not applicable. Table \ref{table:truncated}'s fourth numeric column shows the $MAE$ (Mean Average Error) --- a ranking metric which can also be used to measure string similarity. It measures the average distance from each predicted rank position to the solver's index in the ground truth. Finally, 
the fifth numeric column of Table \ref{table:truncated} shows
the mean regression error (\textit{Reg. error}) which measures the mean absolute difference in predicted solver costs to actual values. 

\begin{table}
\caption{Comparing the seven prediction models and three theoretical strategies}
\begin{tabularx}{\textwidth}{@{}lYYYYY@{}}
\toprule
 & \textbf{Time (secs)} &  \textbf{nDCG} &  $ R^2 $  &  \textbf{MAE} &  \textbf{Reg. error} \\
\midrule
\textit{\textbf{Best}}    &     12.63 &  1.00 &   - & 0.00 &        0.00 \\
\textit{\textbf{Random}}     &   19.06 &  0.36 &  - & 2.63 &       50.77 \\
\textit{\textbf{Worst}}  &     30.26 &  0.00 &   - & 4.00 &       94.65 \\
\midrule
Random Forest        &    15.02 &  0.48 &   \textbf{0.28} & 2.08 &       \textbf{38.91} \\
 Random Forest (discretised)    &  \textbf{14.92} &  0.48 &  -0.18 & 2.13 &       39.19 \\
Decision Tree         &    15.80 &  0.50 &   0.11 & 2.06 &       43.12 \\
K-Nearest Neighbours  &  15.93 &  \textbf{0.53} &   0.16 & \textbf{2.00} &       43.41 \\
Support Vector Regressor   & 15.57 &  0.47 &   0.14 & 2.26 &       47.45 \\
Linear Regression       &   15.17 &  0.42 &  -0.16 & 2.45 &       49.25 \\
Ridge      &   15.11 &  0.42 &  -0.15 & 2.45 &       49.09 \\
\bottomrule
\end{tabularx}
\label{table:truncated}
\end{table}

\subsection{Discussion: choosing a prediction model}
An interesting feature of all the best-performing models in Table \ref{table:truncated} is their ability to predict \textit{multi-output} variables \cite{multisurvey}. In contrast to the Support Vector model, for example, which must predict the cost for each solver individually, a multi-output model predicts each solver's cost simultaneously. Not only is this method more efficient (by reducing the number of estimators required), but it has the ability to account for the correlation of the response variables. This is a useful property in the software verification domain where certain goals are not provable and others are trivial for SMT solvers. Multiple versions of the same solver can also be expected to have highly correlated costs.

After inspecting the results for all learning algorithms (summarised in Table \ref{table:truncated}), we can see that random forests \cite{RandomForests} perform well, relative to other methods. They score highest for three of the five metrics (shown in bold) and have generally good scores in the others. Random forests are an ensemble extension of decision trees: random subsets of the training data are used to train each tree. For regression tasks, the set of predictions for each tree is averaged to obtain the forest's prediction. This method is designed to prevent over-fitting. 

Based on the data in Table \ref{table:truncated} we selected random forests as the choice of predictor to use in \textsf{Where4}.

\section{Implementing \textsf{Where4} in OCaml}

\textsf{Where4}'s interaction with \textsf{Why3} is inspired by the use of machine learning in the Sledgehammer tool \cite{Sledgehammer} which allows the use of SMT solvers in the interactive theorem prover Isabelle/HOL. We aspired to Sledgehammer's `zero click, zero maintenance, zero overhead' philosophy in this regard: it should not interfere with a \textsf{Why3} user's normal work-flow nor should it penalise those who do not use it. 

We implement a ``pre-solving'' heuristic commonly used by portfolio solvers \cite{sunny-cp}\cite{Satzilla}: a single solver is called with a short time limit before feature extraction and solver rank prediction takes place. 
By using a good ``pre-solver'' at this initial stage, easily-proved instances are filtered with a minimum time overhead. We used a ranking of solvers based on the number of goals each could prove, using the data from Table \ref{table:avgtimes}.
The highest-ranking solver installed locally is chosen as a pre-solver. For the purposes of this paper which assumes all 8 solvers are installed, the pre-solver corresponds to Alt-Ergo version 1.01.
The effect pre-solving has on the method \textsf{Where4} uses to return responses is illustrated in Alg. \ref{algo:where4}.      

The random forest is fitted on the entire training set and encoded as a JSON file for legibility and modularity. This approach allows new trees and forests devised by the user (possibly using new SMT solvers or data) to replace our model.  When the user installs \textsf{Where4} locally, this JSON file is read and printed as an OCaml array. For efficiency, other important configuration information is compiled into OCaml data structures at this stage: e.g. the user's \texttt{why3.conf} file is read to determine the supported SMT solvers. All files are compiled and a native binary is produced. This only needs to be done once (unless the locally installed provers have changed). 

The \textsf{Where4} command-line tool has the following functionality:
\begin{enumerate}
\item Read in the WhyML/Why file and extract feature vectors from its goals.
\item Find the predicted costs for each of the 8 provers by traversing the random forest, using each goal's feature vector.
\item Sort the costs to produce a ranking of the SMT solvers.
\item Return a predicted ranking for each goal in the file, without calling any solver .
\item Alternatively, use the \textsf{Why3} API to call each solver (if it is installed) in rank order until a \textit{Valid/Invalid} answer is returned (using Alg. \ref{algo:where4}).

\end{enumerate}

\begin{algorithm}
	\caption{Returning an answer and runtime from a \why~input program}
	\KwIn{$P$, a \why~program; \\ 
		$R$, a static ranking of solvers for pre-proving; \\
		$\phi$, a timeout value}
	\KwOut{$\langle A,T\rangle$ where\\$A$ = the best answer from the solvers;\\
		$T$ = the cumulative time taken to return $A$}
	\Begin{
		\tcc{Highest ranking solver installed locally}
		$S \leftarrow BestInstalled(R) $ \\			
		\tcc{Call solver $S$ on \why~program $P$ with a timeout of 1 second}	
		$\langle A,T \rangle \leftarrow Call(P, S, 1)$ \\
		\If{$A \in \lbrace Valid, Invalid \rbrace $}
		{\Return{$\langle A,T \rangle$}}
		\tcc{extract feature vector $F$ from program $P$} 
		$F \leftarrow ExtractFeatures(P) $ \\
		\tcc{$R$ is now based on program features}
		$R \leftarrow PredictRanking(F) $ \\		
		\While{$A \notin \lbrace Valid, Invalid \rbrace \wedge R \neq \emptyset$}
		{
			$S \leftarrow BestInstalled(R) $ \\	
			\tcc{Call solver $S$ on \why~program $P$ with a timeout of $\phi$ seconds}	
			$\langle A_S,T_S \rangle \leftarrow Call(P, S, \phi)$ \\
			\tcc{add time $T_S$ to the cumulative runtime}
			$T \leftarrow T + T_S$  \\
			\If{$A_S > A$}
			{
				\tcc{answer $A_S$ is better than the current best answer}		
				$A \leftarrow A_S$ }
			\tcc{remove $S$ from the set of solvers $R$}
			$R \leftarrow R \setminus \lbrace S \rbrace$}
		\Return{$\langle A,T\rangle$}} 
	\label{algo:where4}
	
\end{algorithm}

If the user has selected that \textsf{Where4} be available for use through \textsf{Why3}, the file which lets \textsf{Why3} know about supported provers installed locally is modified to contain a new entry for the \textsf{Where4} binary. A simple driver file (which just tells \textsf{Why3} to use the Why logical language for encoding) is added to the drivers' directory. At this point, \textsf{Where4} can be detected by \textsf{Why3}, and then used at the command line, through the IDE or by the OCaml API just like any other supported solver. 

\section{Evaluating \textsf{Where4}'s performance on test data}
\label{sec:eval}

The evaluation of \textsf{Where4} was carried out on a test set of 32 WhyML files, 77 theories, 263 goals (representing 25\% of the entire dataset). This section is guided by the following three Evaluation Criteria:

\begin{table}
\caption{Number of files, theories and goals proved by each strategy and individual solver. The percentage this represents of the total 32 files, 77 theories and 263 goals and the average time (in seconds) are also shown.}
\begin{tabularx}{\textwidth}{@{}l|YYY|YYY|YYY@{}}
\toprule
{} & \multicolumn{3}{c|}{\textbf{File}} & \multicolumn{3}{c|}{\textbf{Theory}} & \multicolumn{3}{c}{\textbf{Goal}} \\
{} & \# proved & \% proved & Avg time & \# proved & \% proved & Avg time & \# proved & \% proved & Avg time \\
\midrule
\textbf{\textsf{Where4}} & 11 & 34.4\% & 1.39 &  44 & 57.1\% & 0.99 & 203 & 77.2\% & 1.98 \\
\textit{Best} & \downbar  & \downbar & 0.25 & \downbar & \downbar & 0.28 & \downbar & \downbar & 0.37 \\
\textit{Random} & \downbar & \downbar & 4.19 & \downbar & \downbar & 4.02 & \downbar & \downbar & 5.70 \\
\textit{Worst} & \upbar & \upbar & 14.71 & \upbar & \upbar & 13.58 & \upbar & \upbar & 18.35 \\
\midrule
\textbf{Alt-Ergo-0.95.2} & 8 & 25.0\% & 0.78 & 37 & 48.1\%& 0.26 & 164 & 62.4\% & 0.34 \\ 
\textbf{Alt-Ergo-1.01} & 10 & 31.3\% & 1.07 & 39 & 50.6\% & 0.26 & 177 & 67.3\% & 0.33 \\ 
\textbf{CVC3} & 5 & 15.6\% & 0.39 & 36 & 46.8\% & 0.21 & 167 & 63.5\% & 0.38 \\ 
\textbf{CVC4} & 4  & 12.5\% & 0.56 & 32 & 41.6\% & 0.21 & 147 & 55.9\% & 0.35 \\ 
\textbf{veriT} & 2 & 6.3\% & 0.12 & 24 & 31.2\% & 0.12 & 100 & 38.0\% & 0.27 \\ 
\textbf{Yices} & 4 & 12.5\% & 0.32 & 32 & 41.6\% & 0.15 & 113 & 43.0\% & 0.18 \\ 
\textbf{Z3-4.3.2} & 6 & 18.8\% & 0.46 & 31 & 40.3\% & 0.20 & 145 & 55.1\% & 0.37 \\ 
\textbf{Z3-4.4.1} & 6 & 18.8\% & 0.56 & 31 & 40.3\% & 0.23 & 145 & 55.1\% & 0.38 \\ 
\bottomrule
\end{tabularx}
\label{table:avgtimes2}
\end{table}

\subsection{EC1: How does \textsf{Where4} perform in comparison to the 8 SMT solvers under consideration?}

\begin{figure}
	\centering
	\includegraphics[width=0.9\linewidth]{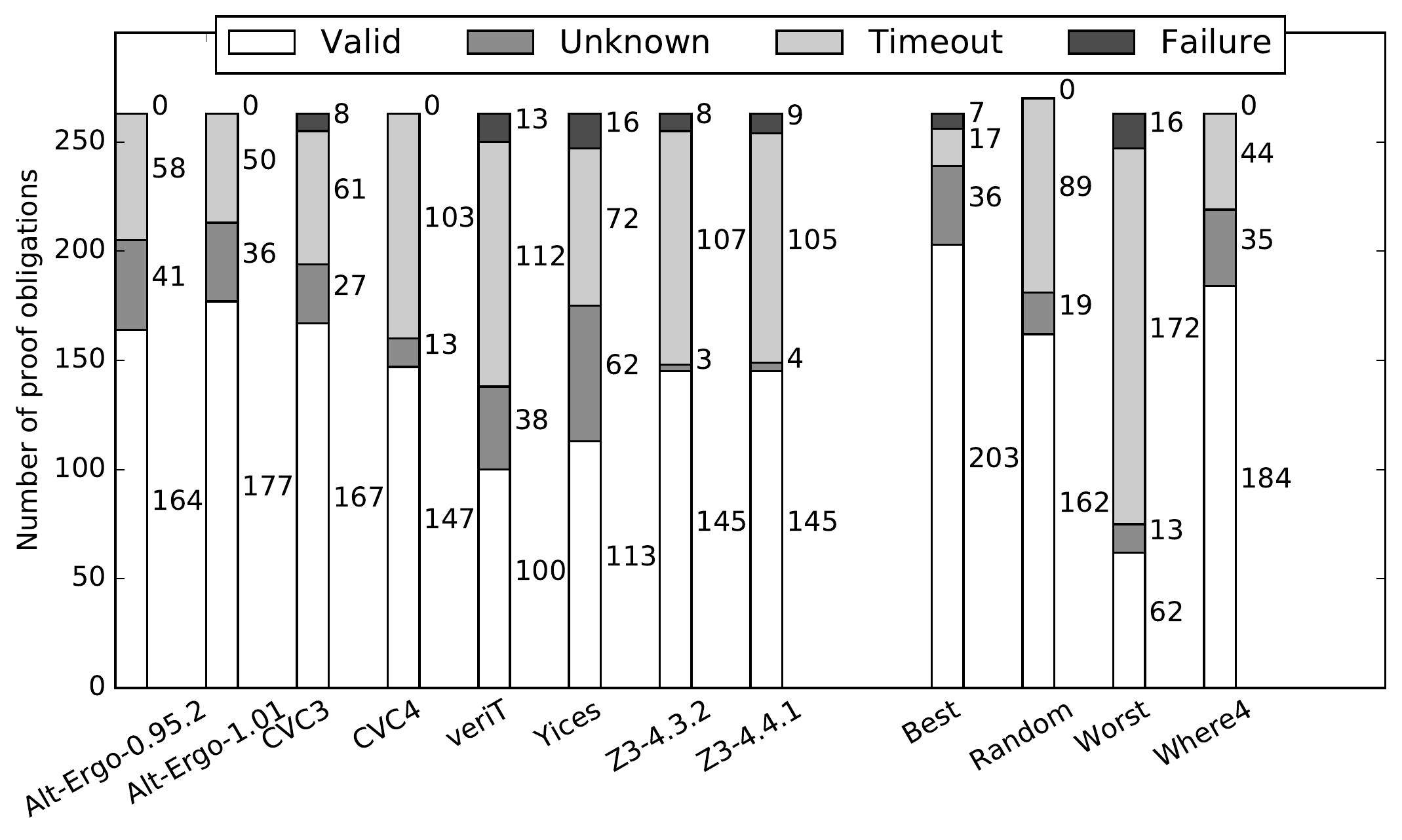}
	\caption{The relative amount of Valid/Unknown/Timeout/Failure answers from the eight SMT solvers. Shown on the right are results obtainable by using the top solver (only) with the 3 ranking strategies and the \textsf{Where4} predicted ranking (with an Alt-Ergo-1.01 pre-solver).}
	\label{fig:barchart2}
\end{figure}

\begin{figure}
	\centering
	\includegraphics[width=0.8\linewidth]{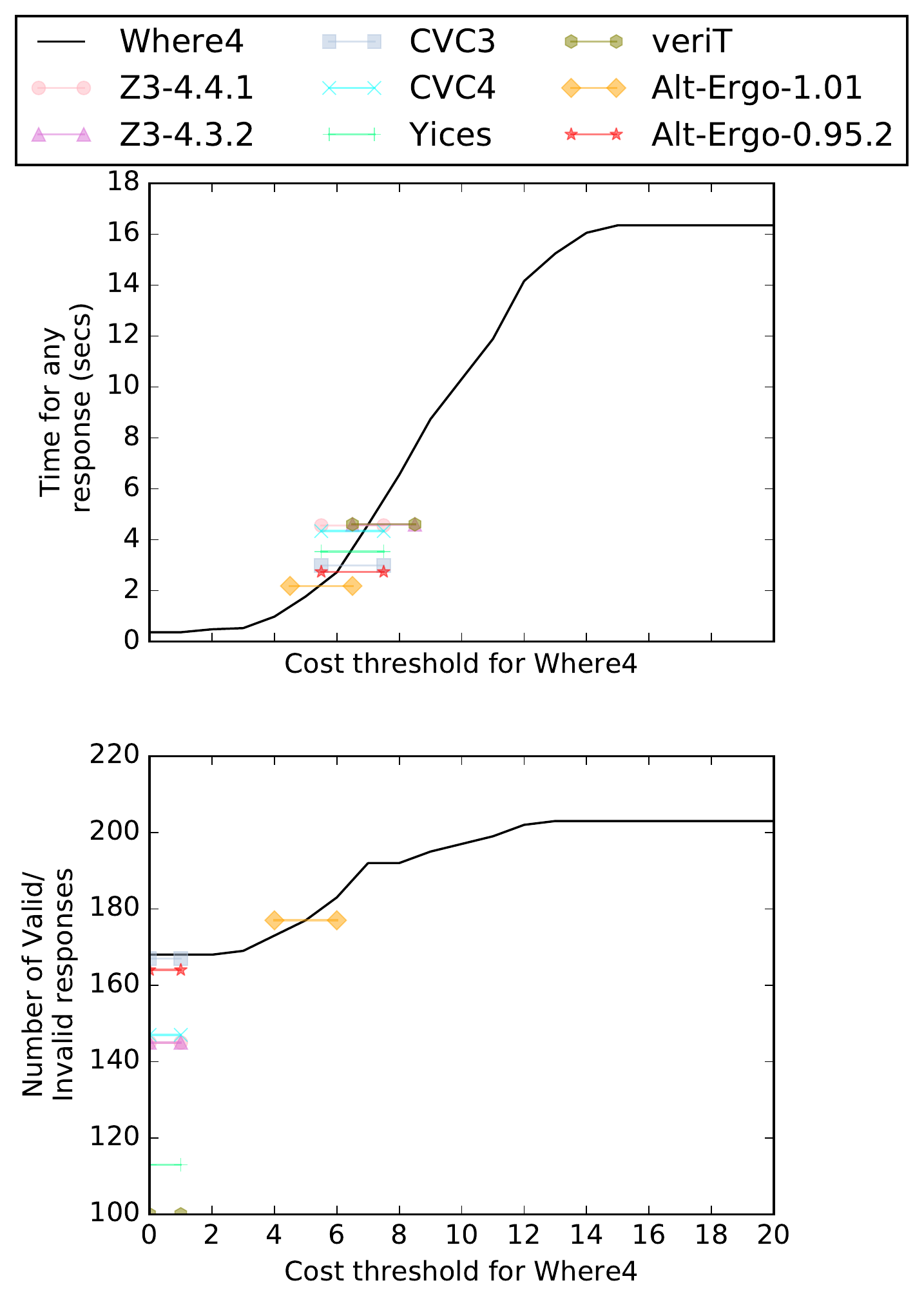}
	\caption{The effect of using a cost threshold. (\textit{top}) The average time taken for \textsf{Where4} to return an answer compared to 8 SMT solvers. (\textit{bottom}) The number of Valid/Invalid answers returned by \textsf{Where4} compared to 8 SMT solvers. For the 7 solvers other than Alt-Ergo-1.01, the number of provable goals is indicated by a mark on the y-axis rather than an intersection with \textsf{Where4}'s results.}
	\label{fig:thresholds}
\end{figure}

\begin{figure}
	\centering
	\includegraphics[width=0.8\linewidth]{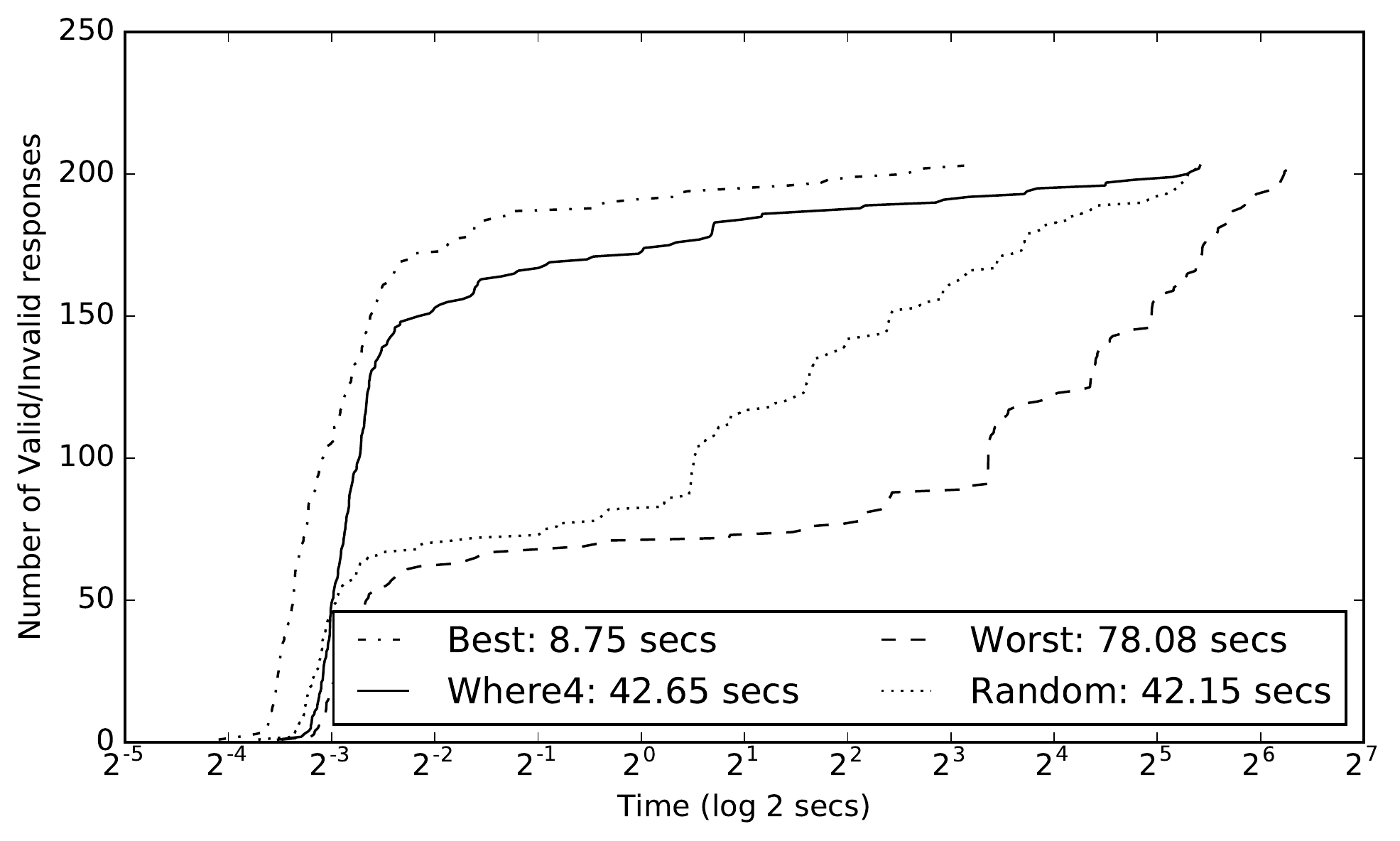}
	\caption{The cumulative time each theoretical strategy, and \textsf{Where4} to return all \textit{Valid/Invalid} answers in the test dataset of 263 goals}
	\label{fig:line-graph-eval-provers}
\end{figure}

When each solver in \textsf{Where4}'s ranking sequence is run on each goal, the maximum amount of files, theories and goals are provable. As Table \ref{table:avgtimes2} shows, the difference between \textsf{Where4} and our set of reference theoretical strategies (\textit{Best, Random}, and \textit{Worst}) is the amount of time taken to return the \textit{Valid/Invalid} result. Compared to the 8 SMT provers, the biggest increase is on individual goals: \textsf{Where4} can prove 203 goals, which is 26 (9.9\%) more goals than the next best single SMT solver, Alt-Ergo-1.01.

Unfortunately, the average time taken to solve each of these goals is high when compared to the 8 SMT provers. This tells us that \textsf{Where4} can perform badly with goals which are not provable by many SMT solvers: expensive \textit{Timeout} results are chosen before the \textit{Valid} result is eventually returned.   In the worst case, \textsf{Where4} may try and time-out for all 8 solvers in sequence, whereas each individual solver does this just once.   Thus, while having access to more solvers allows more goals to be proved, there is also a time penalty to portfolio-based solvers in these circumstances.

At the other extreme, we could limit the portfolio solver to just using the best predicted individual solver (after ``pre-solving''), eliminating the multiple time-out overhead.
Fig. \ref{fig:barchart2} shows that the effect of this is to reduce the number of goals provable by \textsf{Where4}, though this is still more than the best-performing individual SMT solver, Alt-Ergo-1.01.

To calibrate this cost of \textsf{Where4} against the individual SMT solvers, we introduce the notion of a \textit{cost threshold}: using this strategy, after pre-solving, solvers with a predicted cost above this threshold are not called. If no solver's cost is predicted below the threshold, the pre-solver's result is returned.  

Fig. \ref{fig:thresholds} shows the effect of varying this threshold, expressed in terms of the average execution time (top graph) and the number of goals solved (bottom graph).  As we can see from both graphs in Fig. \ref{fig:thresholds}, for the goals in the test set a threshold of 7 for the cost function allows \textsf{Where4} to prove more goals than any single solver, in a time approximately equal to the four slower solvers (CVC4, veriT and both versions of Z3). 

\subsection{EC2: How does \textsf{Where4} perform in comparison to the 3 theoretical ranking strategies?}
Fig. \ref{fig:line-graph-eval-provers} compares the cumulative time taken for \textsf{Where4} and the 3 ranking strategies to return the 203 valid answers in the test set. Although both \textsf{Where4} and \textit{Random} finish at approximately the same time, \textsf{Where4} is significantly faster for returning \textit{Valid/Invalid} answers. \textsf{Where4}'s solid line is more closely correlated to \textit{Best}'s rate of success than the erratic rate of the \textit{Random} strategy. \textit{Best}'s time result shows the capability of a perfect-scoring learning strategy. It is motivation to further improve \textsf{Where4} in the future.   

\subsection{EC3: What is the time overhead of using \textsf{Where4} to prove \textsf{Why3} goals?}The timings for \textsf{Where4} in all plots and tables are based solely on the performance of the constituent solvers (the measurement of which is discussed in Sec. \ref{sec:dependant}). They do not measure the time it takes for the OCaml binary to extract the static metrics, traverse the decision trees and predict the ranking. We have found that this adds (on average) 0.46 seconds to the time \textsf{Where4} takes to return a result for each file. On a per goal basis, this is equivalent to an increase in 0.056 seconds. 

The imitation of an orthodox solver to interact with \textsf{Why3} is more costly: this is due to \textsf{Why3} printing each goal as a temporary file to be read in by the solver individually. Future work will look at bypassing this step for WhyML files while still allowing files to be proved on an individual theory and goal basis.    

\subsection{Threats to Validity}

We categorise threats as either \textit{internal} or \textit{external}.
Internal threats refer to influences that can affect the response variable without the researcher's knowledge and threaten the conclusions reached about the \textit{cause} of the experimental results \cite{experimentation}. 
Threats to external validity are conditions that limit the generalisability and reproducibility of an experiment. 

\subsubsection{Internal}
\label{sec:internal}

The main threat to our work's internal validity is selection bias. All of our training and test samples are taken from the same source. We took care to split the data for training and testing purposes on a \textit{per file} basis. This ensured that \textsf{Where4} was not trained on a goal belonging to the same theory or file as any goal used for testing. 
The results of running the solvers on our dataset are imbalanced. There were far more \textit{Valid} responses than any other response. No goal in our dataset returned an answer of \textit{Invalid} on any of the 8 solvers. This is a serious problem as \textsf{Where4} would not be able to recognize such a goal in real-world use. In future work we hope to use the TPTP benchmark library to remedy these issues. The benchmarks in this library come from a diverse range of contributors working in numerous problem domains \cite{Sutcliffe200139} and are not as specific to software verification as the \textsf{Why3} suite of examples.

Use of an independent dataset is likely to influence the performance of the solvers. Alt-Ergo was designed for use with the \textsf{Why3} platform --- its input language is a previous version of the Why logic language. It is natural that the developers of the \textsf{Why3} examples would write programs which Alt-Ergo in particular would be able to prove. Due to the syntactic similarities in input format and logical similarities such as support for type polymorphism, it is likely that Alt-Ergo would perform well with any \textsf{Why3} dataset. We would hope, however, that the gulf between it and other solvers would narrow.

There may be confounding effects in a solver's results that are not related to the independent variables we used (Sec. \ref{sec:independant}). We were limited in the tools available to extract features from the domain-specific Why logic language (in contrast to related work on model checkers which use the general-purpose C language \cite{DPVZ15:CAV}\cite{MUX}). We made the decision to keep the choice of independent variables simple in order to increase generalisability to other formalisms such as Microsoft's Boogie \cite{Boogie} intermediate language.  

\subsubsection{External}

The generalisability of our results is limited by the fact that all dependent variables were measured on a single machine.\footnote{All data collection was conducted on a 64-bit machine running Ubuntu 14.04 with a dual-core Intel i5-4250U CPU and 16GB of RAM.} We believe that the number of each response for each solver would not vary dramatically on a different machine of similar specifications. By inspecting the results when each solver was given a timeout of 60 seconds (Fig. \ref{fig:line-graph}), the rate of increase for \textit{Valid/Invalid} results was much lower than that of \textit{Unknown/Failure} results. The former set of results are more important when computing the cost value for each solver-goal pair.

Timings of individual goals are likely to vary widely (even across independent executions on the same machine).
It is our assumption that although the actual timed values would be quite different on any other machine, the \textit{ranking} of their timings would stay relatively stable.

A ``typical'' software development scenario might involve a user verifying a single file with a small number of resultant goals: certainly much smaller than the size of our test set (263 goals). In such a setting, the productivity gains associated with using \textsf{Where4} would be minor. \textsf{Where4} is more suited therefore to large-scale software verification.

\subsection{Discussion}
\label{sec:eval-discuss}

By considering the answers to our three Evaluation Criteria, we can make assertions about the success of \textsf{Where4}.
The answer to EC1, \textsf{Where4}'s performance in comparison to individual SMT solvers, is positive.
A small improvement in \textit{Valid/Invalid} responses results from using only the top-ranked solver, while a much bigger increase can be seen by making the full ranking of solvers available for use.
The time penalty associated with calling a number of solvers on an un-provable proof obligation is mitigated by the use of a \textit{cost threshold}.
Judicious use of this threshold value can balance the time-taken-versus-goals-proved trade-off: in our test set of 263 POs, using a threshold value of 7 results in 192 \textit{Valid} responses -- an increase of 15 over the single best solver -- in a reasonable average time per PO (both \textit{Valid} and otherwise) of 4.59 seconds.

There is also cause for optimism in \textsf{Where4}'s performance as compared to the three theoretical ranking strategies --- the subject of Evaluation Criterion 2. 
All but the most stubborn of \textit{Valid} answers are returned in a time far better than \textit{Random} theoretical strategy.
We take this random strategy as representing the behaviour of the non-expert \textsf{Why3} user who does not have a preference amongst the variety of supported SMT solvers.
For this user, \textsf{Where4} could be a valuable tool in the efficient initial verification of proof obligations through the \textsf{Why3} system.    

In terms of time overhead --- the concern of EC3 --- our results are less favourable, particularly when \textsf{Where4} is used as an integrated part of the \textsf{Why3} toolchain.
The costly printing and parsing of goals slows \textsf{Where4} beyond the time overhead associated with feature extraction and prediction.
At present, due to the diversity of languages and input formats used by software verification tools, this is an unavoidable pre-processing step enforced by \textsf{Why3} (and is indeed one of the \textsf{Why3} system's major advantages).

Overall, we believe that the results for two out of three Evaluation Criteria are encouraging and suggest a number of directions for future work to improve \textsf{Where4}.

\section{Comparison with Related Work}
\label{sec:related}
\textit{Comparing verification systems:} The need for a standard set of benchmarks for the diverse range of software systems is a recurring issue in the literature \cite{Dagstuhl}. The benefits of such a benchmark suite are identified by the SMTLIB \cite{SMTLIB} project. The performance of SMT solvers has significantly improved in recent years due in part to the standardisation of an input language and the use of standard benchmark programs in  competitions \cite{SMTEVAL2013}\cite{SVCOMP}. The TPTP (Thousands of Problems for Theorem Provers) project \cite{TPTP} has similar aims but a wider scope: targeting theorem provers which specialise in numerical problems as well as general-purpose SAT and SMT solvers. The TPTP library is specifically designed for the rigorous experimental comparison of solvers \cite{Sutcliffe200139}.           

\textit{Portfolio solvers:} Portfolio-solving approaches have been implemented successfully in the SAT domain by SATzilla \cite{Satzilla} and the constraint satisfaction / optimisation community by tools such as CPHydra \cite{CPHydra} and sunny-cp \cite{sunny-cp}. Numerous studies have used the SVCOMP \cite{SVCOMP} benchmark suite of C programs for model checkers to train portfolio solvers \cite{MUX}\cite{DPVZ15:CAV}. These particular studies have been predicated on the use of Support Vector Machines (SVM) with only a cursory use of linear regression \cite{MUX}. In this respect, our project represents a more wide-ranging treatment of the various prediction models available for portfolio solving. The need for a strategy to delegate \textsf{Why3} goals to appropriate SMT solvers is stated in recent work looking at verification systems on cloud infrastructures \cite{rodinplugin}.

\textit{Machine Learning in Formal Methods:} 
The FlySpec \cite{Flyspec} corpus of proofs has been the basis for a growing number of tools integrating interactive theorem provers with machine-learning based fact-selection. The MaSh engine in Sledgehammer \cite{Sledgehammer} is a related example. It uses a Naive Bayes algorithm and clustering to select facts based on syntactic similarity. 
Unlike \textsf{Where4}, MaSh uses a number of metrics to measure the \textit{shape} of goal formul\ae as features. 
The weighting of features uses an inverse document frequency (IDF) algorithm.
ML4PG (Machine Learning for Proof General) \cite{ML4PG} also uses clustering techniques to guide the user for interactive theorem proving.

Our work adds to the literature by applying a portfolio-solving approach to SMT solvers. We conduct a wider comparison of learning algorithms than other studies which mostly use either SVMs or clustering. Unlike the interactive theorem proving tools mentioned above, \textsf{Where4} is specifically suited to software verification through its integration with the \textsf{Why3} system.

\section{Conclusion and Future Work}

We have presented a strategy to choose appropriate SMT solvers based on \textsf{Why3} syntactic features. Users without any knowledge of SMT solvers can prove a greater number of goals in a shorter amount of time by delegating to \textsf{Where4} than by choosing solvers at random. Although some of \textsf{Where4}'s results are disappointing, we believe that the \textsf{Why3} platform has great potential for machine-learning based portfolio-solving. We are encouraged by the performance of a theoretical \textit{Best} strategy and the convenience that such a tool would give \textsf{Why3} users.

The number of potential directions for this work is large: parallel solving, minimal datasets for practical local training, larger and more generic datasets for increased generalisability, etc. 
The TPTP repository represents a large source of proof obligations which can be translated into the Why logic language. The number of goals provable by \textsf{Where4} could be increased by identifying which goals need to be simplified in order to be tractable for an SMT solver. Splitting transforms would also increase the number of goals for training data: from 1048 to 7489 through the use of the \texttt{split\_goal\_wp} transform, for example.  An interesting direction for this work could be the identification of the appropriate transformations. Also, we will continue to improve the efficiency of \textsf{Where4} when used as a \textsf{Why3} solver and investigate the use of a minimal benchmark suite which can be used to train the model using new SMT solvers and theorem provers installed locally.  

Data related to this paper is hosted at \texttt{github.com/ahealy19/F-IDE-2016}. \textsf{Where4} is hosted at \texttt{github.com/ahealy19/where4}. 

\subsubsection*{Acknowledgments.}This project is being carried out with funding provided by Science Foundation Ireland under grant number 11/RFP.1/CMS/3068

\bibliographystyle{eptcs}
\bibliography{predicting}

\end{document}